# A Reply to Hofman On: "Why LP cannot solve large instances of NP-complete problems in polynomial time"

Moustapha Diaby; January 2, *2007*

*Abstract*— Using an approach that seems to be patterned after that of Yannakakis, Hofman argues that an NP-complete problem cannot be formulated as a polynomial bounded-sized linear programming problem. He then goes on to propose a "construct" that he claims to be a "counter-example" to recently published linear programming formulations of the Traveling Salesman Problem (TSP) and the Quadratic Assignment Problem (QAP), respectively. In this paper we show that Hofman's approach is flawed, and provide further proof that his "counter-example" is invalid.

## I. INTRODUCTION

In [7] Hofman uses an approach that seems to pattern itself after that of Yannakakis [8]. However, there are important differences between Hofman's and Yannakakis' approaches.

***From a dual formulation perspective*** (in order to look at it from the perspective of constraints), Yannakakis [8] shows that the number of constraints required to completely describe the Traveling Salesman Problem (TSP) polytope must be exponential in any linear programming formulation that is *symmetric*. He concludes that the TSP cannot be, therefore, formulated as a "small-sized" linear programming problem over the TSP polytope. (By "small-sized," it is meant a size bounded by a sub-exponential polynomial function of the number of cities).

It is important to note that Yannakakis' [8] approach consists essentially (i.e., *from a dual formulation perspective*) of developing bounds on the number of facets required to describe the polytopes he studies, *based* on the pre-supposition that the (primal) formulation at hand is symmetric. Hence, absent Symmetry in a the primal formulation, the implicit bounds developed in Yannakakis [8] on the number of facets are not valid.

It is important to note also that Yannakakis [8] focuses on specific polytopes (namely, the TSP, (general) matching, and vertex packing polytopes) only.

## II. FLAWS IN HOFMAN'S APPROACH

Unlike Yannakakis [8], Hofman [7] overlooks the fact that the optimization problem involved in any given NP-Complete problem need not be expressed over any particular polytope, *a priori*. Hence, issues of descriptions of *the* specific polytope of a given NP-Complete problem (if such has been defined, as has been done for most of the well-known combinatorial optimization problems) do not automatically apply to any arbitrary mathematical programming model of that NP-Complete problem. For example, this is discussed for the case of the TSP in [2].

Hofman's [7] argumentation is essentially that a polytope with an exponential *number of vertices* cannot be completely described using a polynomially-bounded number of constraints. This is exemplified by the following statements (among many others) in Hofman [7]:

" … *Number of facets is in general exponential for polytopes with exponential number of vertexes.*" (Hofman [7; page 2]).

Or:

" … *Model is then stored as polynomial number of equations so 1) is satisfied as well, but we do not have equation for every pair of vertexes (we said that the number of vertexes on outline path is $O(2^n)$, so the 'lowest possible' number of pairs is also $O(2^n)$, storing them would not fir* (fit?) *in polynomially bounded space)…*

*…There are $O(2^n)$ different pairs what means that every model storing information about polytope in polynomially bounded space has to omit some of them.*" (Hofman [7; page 3]).

Hofman's argumentation is flawed because it directly negates some well-established, basic facts about mathematical programming models:

1. It is well-established that the Simplex procedure is not a polynomial-time procedure (see [1]). The reason for this is that the number of extreme-points of a Linear Program tends to be exponential in general;

2. It is inded possible to completely describe a polytope with an exponential number of extreme points using a polynomially-bounded number of constraints. An example of this is the standard assignment polytope (see [1], or [5]);

3. The Linear Programming relaxation of an Integer Program can, under certain conditions (such as total-unimodularity for example), have extreme points that are integral. The standard assignment problem is an example of this also.

## III. HOFMAN'S "COUNTER-EXAMPLE" IS INVALID

Hofman's [6], [7] claim of a counter-example is based on his belief in having "constructed" a solution to a 32-city TSP that includes all the variables and satisfies all the constraints of the LP model in [2] (and [3]) and yet has a lower cost than that of an optimal TSP tour.

In [4], we showed the impossibility for Hofman's "solution" in [6] to be feasible for the models in [2] and [3]. In addition to this, we will point out here that Hofman [6], [7] did not actually consider all the pertinent variables and constraints of the models in [2] and [3].

In [6], Hofman states that:

" *...Program in PERL language generating all necessary variables and every equation described in latest version of article is listed in section 6.2. It will generate 2 files – 33 MB of variables (1,120,335 of them), and 169 MB of equations (997,419 of them). Program checks every equation if it is correct, and of course every equation is. ...*" (Hofman [6; page 24]);

Hofman also states in [7] that:

"… *for 32 nodes, there are almost $10^6$ equations containing non-zero variables.*" (Hofman [7; page 3]).

We point out here that, leaving out variables with zero values and the corresponding constraints, the numbers of variables and explicit constraints called for in the models in [2] (and [3]) are in the hundreds of millions, respectively, for a 32-city TSP (see Table 1).

Hence, in checking the feasibility of his "solution" for the model in [2] (and [3]), Hofman [6], [7] considered only very small fractions of the relevant variables and constraints called for in the model, respectively.

Hence, Hofman's claim of having constructed a "counter-example" is incorrect.

| Number of cities | Number of variables | Number of constraints |
|---|---|---|
| 7 | 8,910 | 8,881 |
| 8 | 63,462 | 40,321 |
| 9 | 372,008 | 141,793 |
| 10 | 1,748,952 | 413,785 |
| 11 | 6,745,050 | 1,049,761 |
| 12 | 22,090,970 | 2,389,861 |

Table 1: *Illustration of the Size of Diaby's LP Model for the TSP*